\documentclass[english]{achemso}
\usepackage[T1]{fontenc}
\usepackage[latin9]{inputenc}
\usepackage{color}
\usepackage{units}
\usepackage{amsmath}
\usepackage{amssymb}
\usepackage{graphicx}

\makeatletter

\title{Graphene Quantum Dot with Divacancy and Topological Defects: A Novel
Material for Promoting Prompt and Delayed Fluorescence of Tunable
Wavelengths }
\author{Tushima Basak}
\affiliation{Department of Physics, \textcolor{black}{SVKM's Mithibai College
of Arts Chauhan Institute of Science \& Amrutben Jivanlal College
of Commerce and Economics}, Mumbai 400056, India}
\email{Tushima.Basak@mithibai.ac.in}
\author{Tista Basak}
\affiliation{Mukesh Patel School of Technology Management and Engineering, NMIMS
University, Mumbai 400056, India}
\email{tista.basak@nmims.edu}
\author{Alok Shukla}
\affiliation{Department of Physics, Indian Institute of Technology Bombay, Powai,
Mumbai 400076, India}
\email{shukla@phy.iitb.ac.in}
\providecommand{\tabularnewline}{\\}
\newcommand{\lyxdot}{.}

\@ifundefined{showcaptionsetup}{}{%
 \PassOptionsToPackage{caption=false}{subfig}}
\usepackage{subfig}
\makeatother

\usepackage{babel}
\begin{document}
\begin{abstract}
\textcolor{black}{This work demonstrates the unique approach of introducing
divacancy imperfections in topological Stone-Wales type defected graphene
quantum dots for harvesting both singlet and triplet excitons, essential
for fabricating fluorescent organic light-emitting diodes. Here, we
first reveal that structural relaxation of these systems establishes
the high-spin triplet state as the stable ground state at room temperature,
thereby significantly increasing their potential in designing spintronic
devices. Our extensive electron-correlated computations then demonstrate
that the energetic ordering of the singlet and triplet states in these
relaxed structures can trigger both prompt and delayed fluorescence
of different wavelengths through various decay channels. Particularly,
the position of divacancy determines the tunability range of the emission
wavelengths. In addition, our results obtained from both multi-reference
singles-doubles configuration-interaction (MRSDCI) and first-principles
time-dependent density functional theory (TDDFT) methodologies highlight
that the synergetic effects of divacancy-position, structural relaxation
and spin multiplicity critically govern the nature and magnitude of
shift exhibited by the most intense peak of the absorption profile,
crucial for designing optoelectronic devices.}
\end{abstract}

\section{Introduction}

\textcolor{black}{The development of unique materials with tunable
electronic and optical properties, required for optimizing the performance
of optoelectronic devices, has been a crucial theoretical and technical
challenge in recent years \cite{Hirata_Phosphorescence}. In principle,
high optical efficiency can be achieved in these new-generation materials
by the simultaneous harvesting of singlet and triplet excitons. In
this work, we propose for the first time, an efficient mechanism of
exploiting the structural relaxation process associated with the suitable
incorporation of divacancy and Stone-Wales (SW) type topological defects
in graphene quantum dots (GQDs) for promoting both prompt and delayed
fluorescence of varied wavelengths, which is crucial for designing
fluorescent organic light-emitting diodes (FOLEDs). }

The formation and reconstruction of \textit{in situ} atomic point
defects (single vacancy and double vacancy) and topological anomalies
like SW-type bond reconstructions, created ubiquitiously during facile
synthesis of graphene, has been precisely imaged by atomic resolution
microscopy techniques like scanning tunneling microscopy (STM) and
aberration corrected transmission electron microscopy (ac-TEM) \cite{Hashimoto_Nature_2004,Meyer_Nano_Letters_2008}.
In particular, TEM measurements \cite{Cretu_PhysRevLett.105.196102}
have demonstrated that double vacancies (DVs) are more abundant than
single vacancies (SVs) at room temperature. Further, ac-TEM measurements
\cite{Girit_Science_2009} have recorded the real-time evolution of
DVs in graphene into different SW-type bond reconstructed patterns
such as triple pentagon-triple heptagon (555-777) and pentagon-octagon-pentagon
(5-8-5) configurations within a time-scale of few seconds.

Theoretical studies have also affirmed the relative stability of various
types of DV reconstructed structures \cite{Kim_PRB_2011} and their
dynamics \cite{Lee_PRL_2005,Leyssale_jp501028n} for a period of few
picoseconds. In case of graphene nanostructures, the absence of dangling
bonds in a DV is responsible for its formation energy being lower
than that of two isolated SVs \cite{Krasheninnikov_2006_CPL,Kotakoski_PRB_2006,El_barbary_PRB_2003}.
Also, the energetic stability of divacancy-defected graphene nanostructures
is determined by the position of the DV \cite{Sahan_PhysicaE_2019},
size and edge termination type (zigzag or armchair) of the nanoflake
\cite{Liu_Hindawi_2017}. The shape of the graphene nanoflakes with
divacancies is also crucial in deciding the spin multiplicity of the
ground state of these structures \cite{Yumura_Molecules_2012}.

Even though a considerable amount of literature is available on the
microscopic imaging, identification of real-time dynamics of carbon
vacancies and their energetics in graphene nanostructures, there is
no report till date to highlight the variation in electronic and optical
properties due to structural relaxation of DVs in SW-defected GQDs.
In this work, we perform extensive electron-correlated computations
based on the Pariser-Parr-Pople (PPP) model Hamiltonian and configuration-interaction
(CI) methodology to systematically explore for the first time, the
imprints of structural relaxation in modulating the electronic and
optical properties of hydrogen-passivated GQDs having divacancy imperfections
as well as armchair, zigzag and topologically SW-reconstructed zigzag
(reczag) edge terminations. \textcolor{black}{In order to compare
and rationalize different computational approaches, we have also employed
the TDDFT methodology to compute the absorption spectra of these systems.
Our computed results demonstrate that structural relaxation, divacancy-position,
and spin multiplicity are pivotal in determining the energetic ordering,
decay channels for prompt as well as delayed fluorescence of tunable
wavelengths and the marked optical absorption spectral features of
singlet and triplet states. }

\section{Theoretical model and computational approach}

\textcolor{black}{The divacancy defect is introduced at (i) the centre
and (ii) the reczag edge of the GQD and their optimized structures
are obtained} by relaxing the configurations using the Gaussian 16
program package \cite{g16}, using a cc-pvdz basis set at restricted
Hartree-Fock (RHF) level. The electron-correlated computations on
these optimized structures are then initiated by applying our theoretical
model based on Pariser-Parr-Pople (PPP) Hamiltonian \cite{ppp-pople,ppp-pariser-parr}
denoted by

\begin{align}
\mbox{\mbox{\mbox{\mbox{\ensuremath{H}}}}} & \ensuremath{=}\ensuremath{-}\ensuremath{\sum_{i,j,\sigma}t_{ij}\left(c_{i\sigma}^{\dagger}c_{j\sigma}+c_{j\sigma}^{\dagger}c_{i\sigma}\right)}\ensuremath{+}\ensuremath{U\sum_{i}n_{i\uparrow}n_{i\downarrow}}\nonumber \\
 & \ensuremath{+}\ensuremath{\sum_{i<j}V_{ij}(n_{i}-1)(n_{j}-1)}\label{eq:ppp-2}
\end{align}

Here, the first term indicates the kinetic energy of the $p_{z}$
electron having spin $\sigma$, when it delocalizes from the \emph{$i^{th}$}
carbon atom to its adjacent $j^{th}$ carbon atom. The parameter $t_{ij}$
represents the hopping matrix term between adjoining carbon atoms
while $c_{i\sigma}^{\dagger}\left(c_{i\sigma}\right)$ creates (annihilates)
the $p_{z}$ electron at $i^{th}$ location. The value of hopping
matrix element, $t_{0},$ is selected as 2.4 eV when the bond length
between adjacent carbon atoms, $R_{0},$ is 1.4 Å. In this work, the
magnitude of hopping matrix elements corresponding to C-C bond lengths
shorter or longer than 1.4 Å are derived from the relationship $t_{ij}=t_{0}e^{(R_{0}-R_{ij})/\delta}$,
with $\delta=0.73$ Å \cite{Tista_PhysRevB.103.235420}. The on-site
Coulomb repulsion between the electrons is denoted by the second term
in Eq. \ref{eq:ppp-2} wherein, $n$$_{i}=\sum_{\sigma}c_{i\sigma}^{\dagger}c_{i\sigma}$
considers all electrons with spin $\sigma$ on \emph{$i^{th}$} carbon
atom. The last term in this equation depicts long-range Coulomb repulsion
with the parameter $V_{ij}$ incorporating screening effects in accordance
to the Ohno relationship \cite{Theor.chim.act.2Ohno}
\begin{equation}
V_{ij}=U/\kappa_{i,j}(1+0.6117R_{i,j}^{2})^{\nicefrac{1}{2}}
\end{equation}

wherein, the screening effects are simulated by the dielectric constant
$\left(\kappa_{i,j}\right)$ of the system and $R_{i,j}$ denotes
the spacing between adjacent carbon atoms (in $\textrm{Å}$). The
screened parameters with $U=8.0$ eV, $\kappa_{i,i}=1.0$ and $\kappa_{i,j}=2.0\left(i\neq j\right)$
are employed, in agreement with our earlier studies \cite{PhysRevB.98.035401,Tista_PRB93,Tista_PRB92,Tista_Basak_Mat_Today_Proc_2020,Tushima_Basak_Mat_Today_Proc_2020}.

For determining the singlet optical absorption spectra of our GQDs,
electron correlated computations are performed by implementing the
MRSDCI methodology. For this objective, a code developed by our group
\cite{Sony2010821} is used to initially transform the PPP Hamiltonian
from the site basis to molecular orbital (MO) basis by employing the
MOs obtained from the mean-field RHF computations. This transformed
hamiltonian and a single reference space comprising of RHF ground
state are considered for calculating the preliminary CI matrix. The
transition electric dipole matrix elements between different excited
states, necessary for calculating the optical absorption spectra,
are then derived from the eigenfunctions generated from this CI matrix.
Thereafter, the excited states leading to peaks in the absorption
profile are discerned and their corresponding reference eigenfunctions
with coefficients above a predetermined convergence criterion are
taken into consideration to create the new reference space for performing
the next MRSDCI calculation. These steps are repeated in chronological
order until convergence of the excitation energies and optical absorption
spectra of the system are achieved within an acceptable tolerance.
In order to make the computations attainable, a small number of initial
low-energy occupied orbitals are frozen and upper high-energy virtual
orbitals are deleted to reduce the set of active MOs.

The transition dipole components and excited state energies are employed
to compute the Lorentzian line shaped ground state optical absorption
cross-section $\sigma(\omega)$ given by,

\begin{equation}
\sigma(\omega)=4\pi\alpha\underset{i}{\sum}\frac{\omega_{i0}\left|\left\langle i\left|\mathbf{\hat{e}.r}\right|0\right\rangle \right|^{2}\gamma}{\left(\omega_{i0}-\omega\right)^{2}+\gamma^{2}},
\end{equation}
 where, $\omega$, $\omega_{i0}$, $\hat{{\bf e}},$ ${\bf r}$, $\alpha$
and $\gamma$ represent the frequency of incident radiation, energy
difference (in frequency units) between ground $\left(\left|0\right\rangle \right)$
and excited $\left(\left|i\right\rangle \right)$ levels, polarization
direction of incident radiation, position operator, fine structure
constant and absorption line-width, respectively. 

For computing the triplet absorption spectrum, the same methodology
is adopted by considering the many-electron wave-functions having
triplet spin multiplicity and then deriving the corresponding transition
dipole elements and excited triplet state energies.

\textcolor{black}{For the purpose of authenticating our PPP-model-based
configuration-interaction (PPP-CI) results, the optical absorption
spectra of the optimized configurations (B3LYP/6-31g(d)) are computed
by employing the TD-B3LYP approach using the Gaussian 16 program package
\cite{g16}. This framework is adopted as it can accurately describe
the ground and excited-state properties of medium to large-sized nanostructures
at a low computational cost. }

\section{Results and Discussion}

\subsection{Structure and energetics}

\begin{figure}
\subfloat[\label{fig:GQD-DV1}]{\includegraphics[scale=0.1]{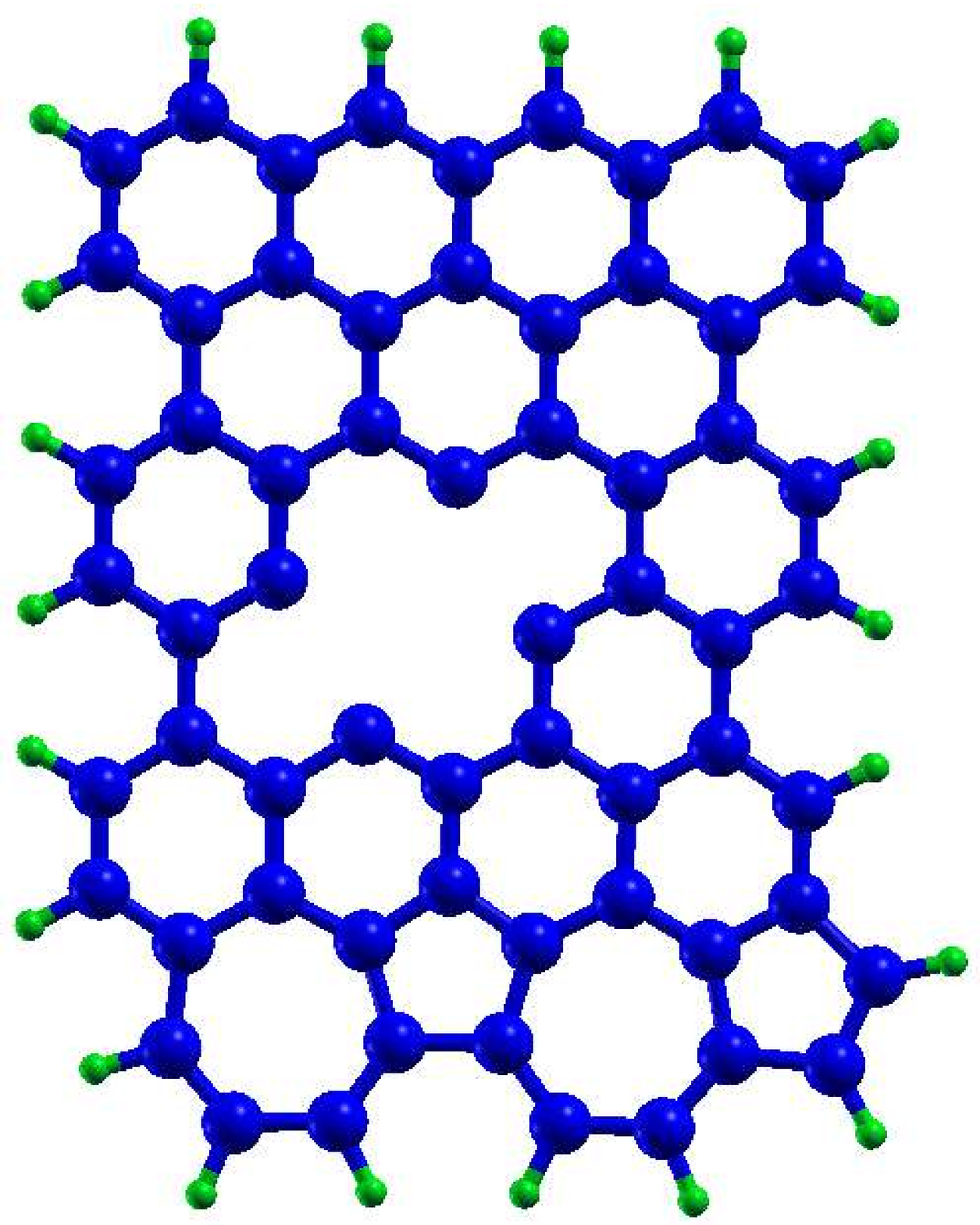}

}\subfloat[\label{fig:GQD-DV2}]{\includegraphics[scale=0.1]{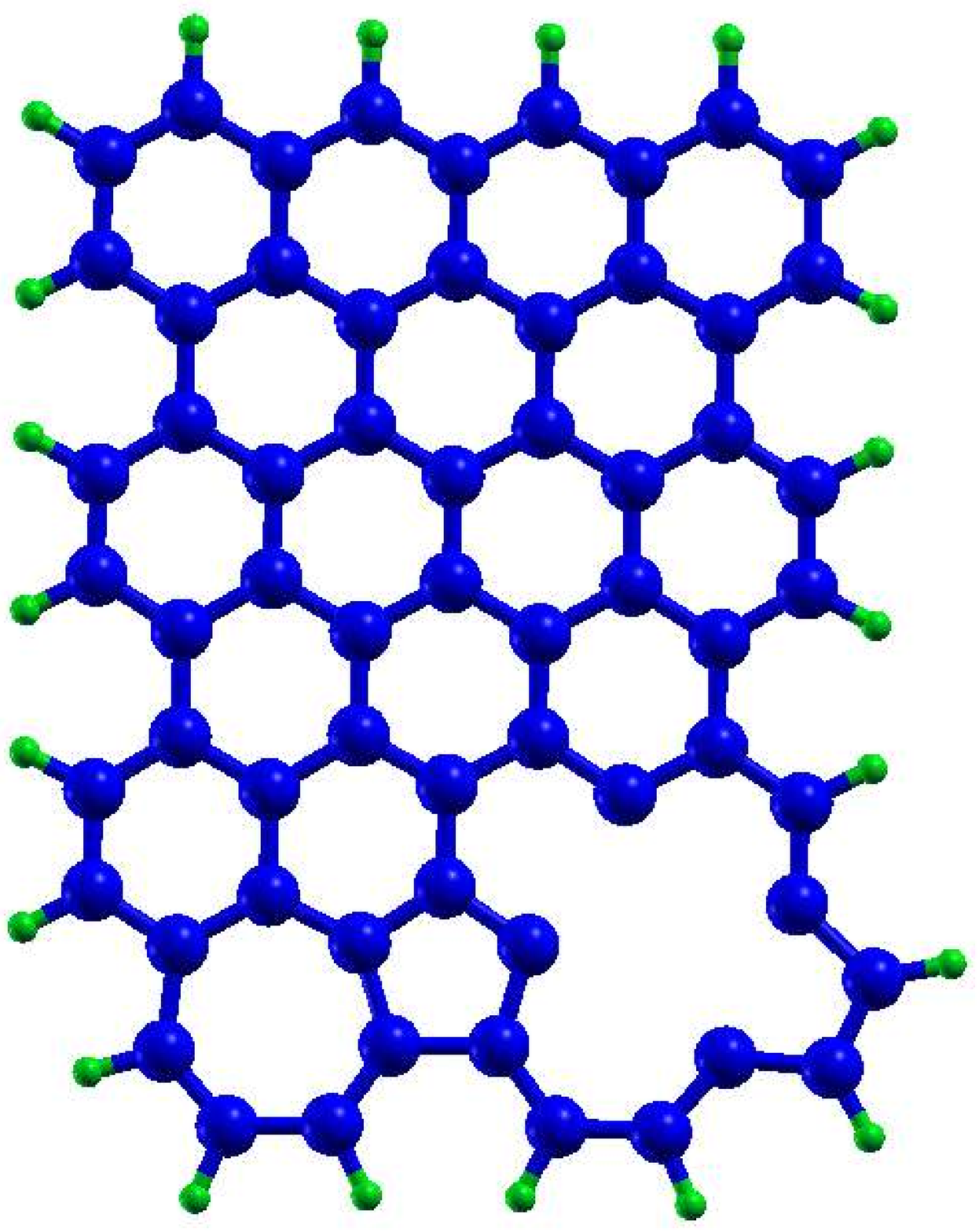}

}\subfloat[\label{fig:GQD-DV1-585}]{\includegraphics[scale=0.1]{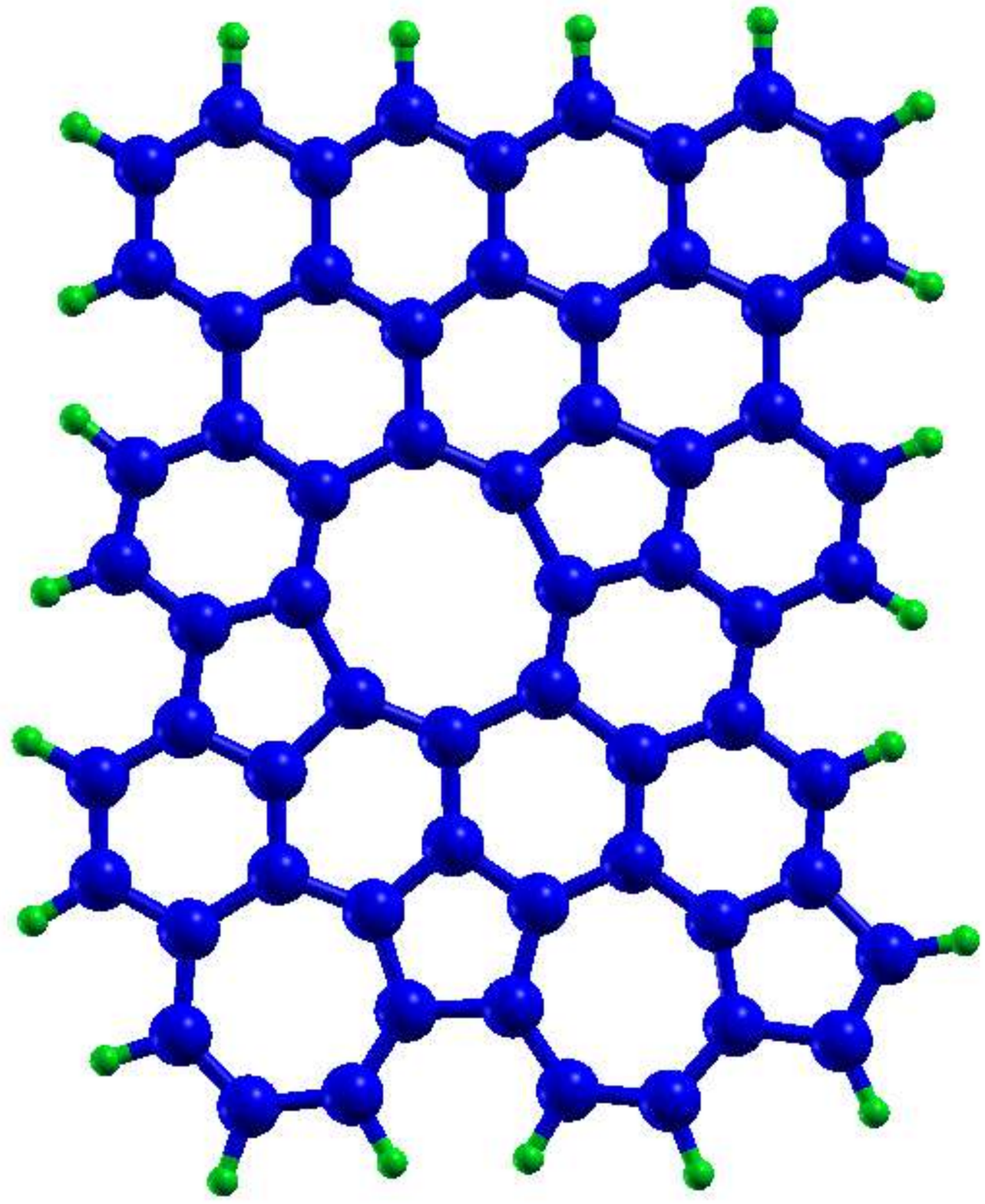}

}\subfloat[\label{fig:GQD-DV2-594}]{\includegraphics[scale=0.1]{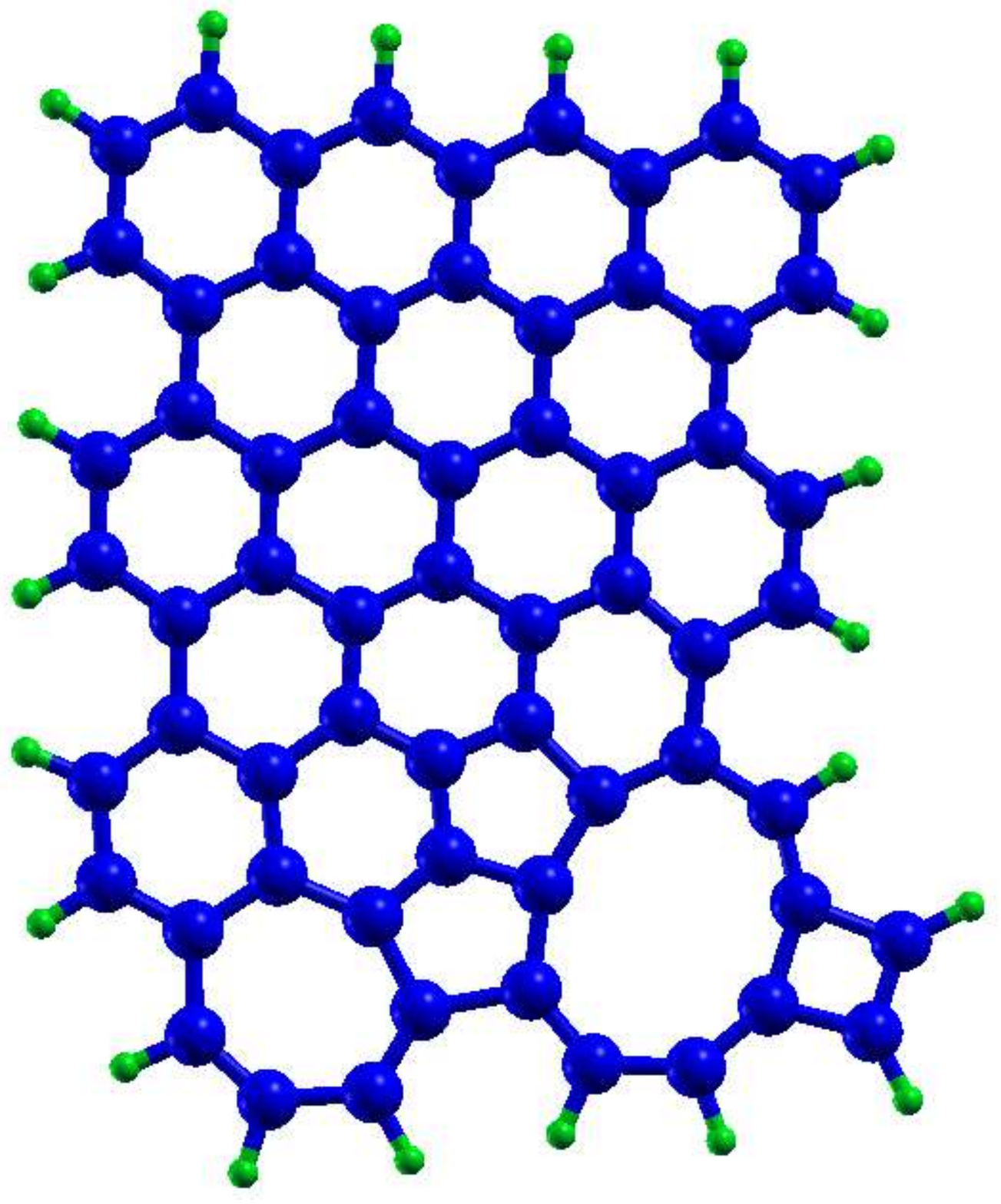}}\caption{Schematic representation of (a) GQD-DV1, (b) GQD-DV2, (c) GQD-DV1-585,
and (d) GQD-DV2-594 systems. \textcolor{black}{The blue and green
spheres denote carbon and hydrogen atoms, respectively. }}
\end{figure}

Figure (\ref{fig:GQD-DV1} - \ref{fig:GQD-DV2}) displays the initial
unrelaxed geometrical configuration of hydrogen-passivated GQDs with
armchair, zigzag and reczag edges having divacancies at the centre
(GQD-DV1) and at the reczag edge (GQD-DV2). The optimized geometries
GQD-DV1-585 and GQD-DV2-594 obtained after structural relaxation of
GQD-DV1 and GQD-DV2 systems are depicted in figs. \ref{fig:GQD-DV1-585}
and \ref{fig:GQD-DV2-594}, respectively. Our results demonstrate
that the geometry of the systems is altered significantly on relaxation.
The divacancies at the centre reconstruct into a pattern composed
of two pentagons and a central octagon while the DVs at the reczag
edge transform into an arrangement consisting of a nonagon in between
a pentagon and a tetragon. The bond-lengths between the adjacent carbon
atoms are in the range (1.35 - 1.49) Å for the unrelaxed systems while
they vary from 1.33 $\textrm{Å}$ to 1.61 $\textrm{Å}$ for the relaxed
configurations. \textcolor{black}{The coordinates of the carbon and
hydrogen atoms in all these systems are listed in Table S1 of the
Supporting Information \cite{Supp_Info}. A detailed analysis of the
structural energetics of optimized GQD-DV1-585 and GQD-DV2-594 geometries
(obtained from Gaussian 16 program package) reveals that they have
equal energies, implying that their structural stability is independent
of the position of divacancies. }

\subsection{Energetic ordering and decay mechanism of triplet and singlet states}

\begin{figure}
\includegraphics[scale=0.4]{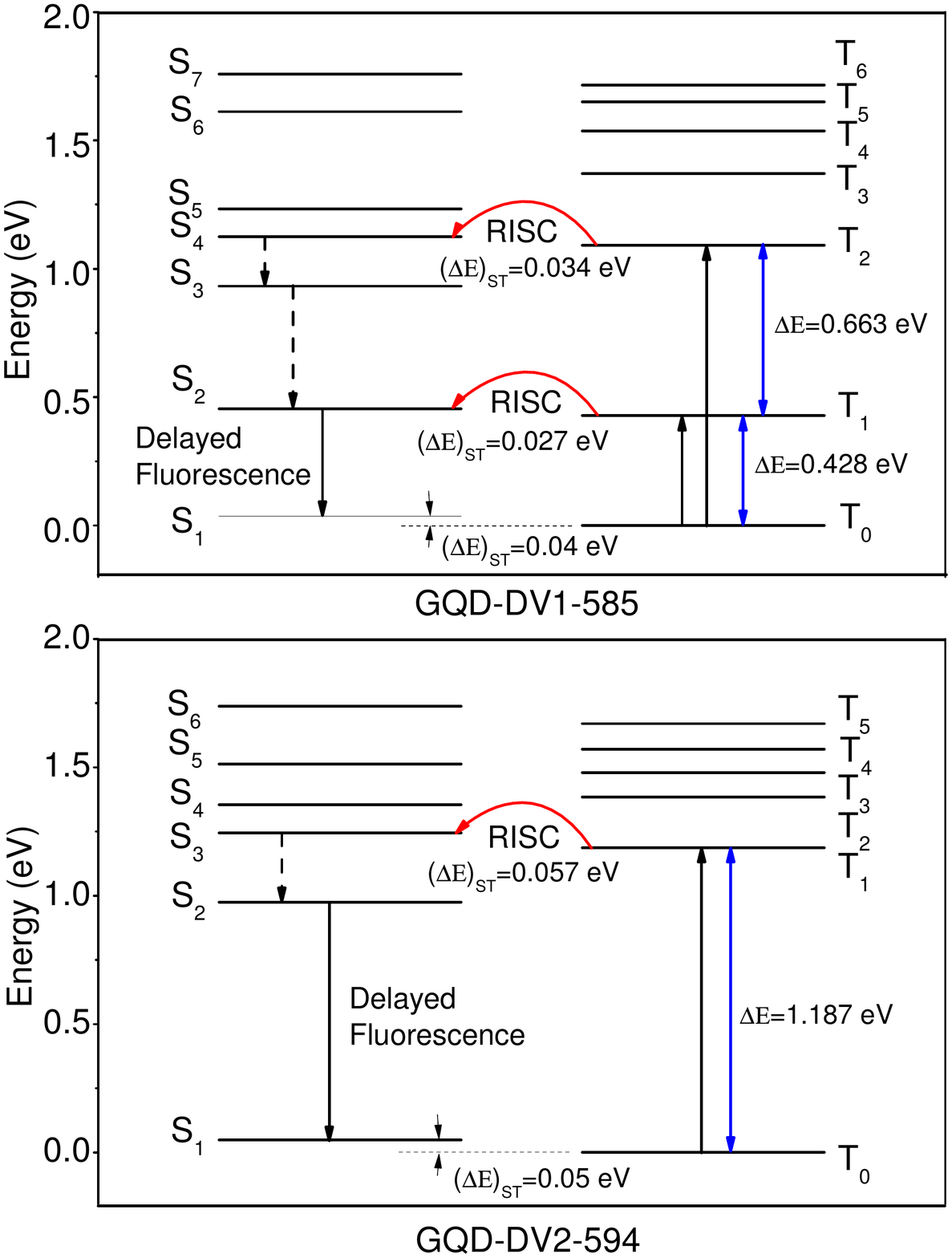}

\caption{Energetic ordering \label{fig:Energetic-ordering-of}of the first
few singlet and triplet states of GQD-DV1-585 and GQD-DV2-594 computed
by adopting PPP-CI approach. }
\end{figure}

Figure \ref{fig:Energetic-ordering-of} exhibits the energetic ordering
of the first few singlet $(S_{m},m\geqslant1)$ and triplet $(T_{n},n\geqslant0)$
states of the relaxed systems considered in this work, computed using
the PPP-CI methodology. Our investigations reveal that the lowest
energy singlet state is below the first triplet state for the unrelaxed
systems, while the trend is reversed for their relaxed counterparts.\textcolor{black}{{}
The energy difference between the first singlet $\left(S_{1}\right)$
and triplet $\left(T_{0}\right)$ states for GQD-DV1-585 and GQD-DV2-594
are 0.04 eV and 0.05 eV, respectively, implying that the high-spin
triplet state will be the stable ground state of these systems at
room temperature (0.025 eV). Hence, the high-spin ground state of
these relaxed configurations having robust thermal stability at room
temperature significantly enhances the potential application of these
GQDs in the field of organic spintronic devices. }

\begin{figure}
\includegraphics[scale=1.3]{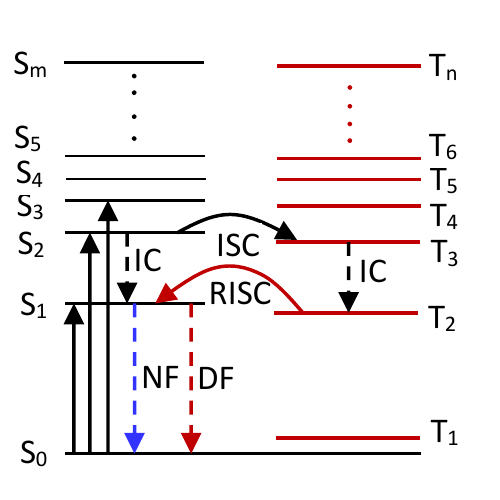}

\caption{\textcolor{black}{Schematic representation of prompt} \label{fig:Schematic-representation-of}\textcolor{black}{or
normal fluorescence (NF), delayed fluorescence (DF), internal conversion
(IC), intersystem crossing (ISC) and reverse intersystem crossing
(RISC) processes occurring when divacancy and SW-defected GQDs are
in the singlet ground state.}}
\end{figure}

\textcolor{black}{The low energy triplet excitons in the $T_{1}$
level of GQD-DV1-585 and GQD-DV2-594 can promptly fluoresce to the
$T_{0}$ state emitting wavelengths (2897 nm) and (1044 nm), respectively,
in the near-infrared (NIR) range. Alternately, the marginal energy
splitting between the upper triplet and singlet excited levels facilitates
the conversion of high-energy triplet excitons to the emissive singlet
level $\left(S_{2}\right)$ via reverse intersystem crossing (RISC)
process. The consequent radiative decay from $S_{2}$ to $S_{1}$
can give rise to delayed fluorescence, with long emission lifetime,
in both these systems. In case of GQD-DV1-585, the small energy difference
between $S_{2}$ and $T_{1}$ ($\Delta E_{ST}=0.027$ eV) as also
between $S_{4}$ and $T_{2}$ ($\Delta E_{ST}=0.034$ eV) states can
lead to delayed fluorescence (2966 nm) along the $T_{1}\rightarrow S_{2}\rightarrow S_{1}$
and $T_{2}\rightarrow S_{4}\rightarrow S_{3}\rightarrow S_{2}\rightarrow S_{1}$
channels, respectively. A transition involving high-lying excited
energy levels $\left(T_{n}\rightarrow S_{m},n\geqslant1,m\geqslant2\right)$
is termed as ``hot exciton'' process. In particular, the large energy
separation between $T_{2}$ and $T_{1}$ (0.663 eV) states substantially
enhances the possibility of RISC ``hot exciton'' process\cite{Hu_RISC}}\textcolor{red}{{}
}\textcolor{black}{in contrast to the internal conversion (IC) of
triplet excitons along $T_{2}\rightarrow T_{1}$ route in this system.
In case of GQD-DV2-594, the very small singlet-triplet energy splitting
($\Delta E_{ST}=0.057$ eV) between the $S_{3}$ and $T_{1}$ states
can prompt delayed fluorescence (1272 nm) along the $T_{1}\rightarrow S_{3}\rightarrow S_{2}\rightarrow S_{1}$
decay path. Hence, our findings illustrate that introducing divacancies
in SW-defected GQDs is an efficient mechanism for realizing both prompt
as well as delayed fluorescence of varied wavelengths and the range
of tunability of these wavelengths is more for GQD-DV2-594 as compared
to GQD-DV1-585. In addition, the energetic proximity of the}\textcolor{red}{{}
}\textcolor{black}{$T_{0}$}\textcolor{red}{{} }\textcolor{black}{and
$S_{1}$ energy levels can induce a spin cross-over of the ground
state of these systems from triplet to singlet at elevated temperatures.
However, even when these systems are in the singlet ground state,
the marginal singlet-triplet energy splitting coupled with the large
energy gaps between the triplet states can trigger delayed fluorescence
via intersystem crossing and RISC processes along with normal luminescence.
A schematic representation of this mechanism is illustrated in fig.}\textcolor{red}{{}
}\textcolor{black}{\ref{fig:Schematic-representation-of}. Hence,
SW-defected GQDs with divacancy can be suitably utilized to enhance
the efficiency of next-generation FOLED devices by harvesting both
triplet and singlet excitons, simultaneously. }

\subsection{Optical properties of divacancy and SW-defected GQDs}

We now investigate the role of structural relaxation and electron-electron
correlation in governing the optical spectral features of all the
above considered divacancy and SW-defected GQDs. \textcolor{black}{For
this purpose, PPP-CI and TDDFT methodologies are adopted to compute
and analyze the linear absorption spectra for both the spin multiplicity
states, since it is feasible for these GQDs to exist in either of
these states at high temperatures.} The precision of our CI results,
in the absence of experimental data, are validated by the large dimensions
($>10^{6}$) of the diagonalized CI matrices (Table S2 \cite{Supp_Info}). 

\begin{figure}
\subfloat[\label{fig:GQD-DV1_Singlet}]{\includegraphics[scale=0.25]{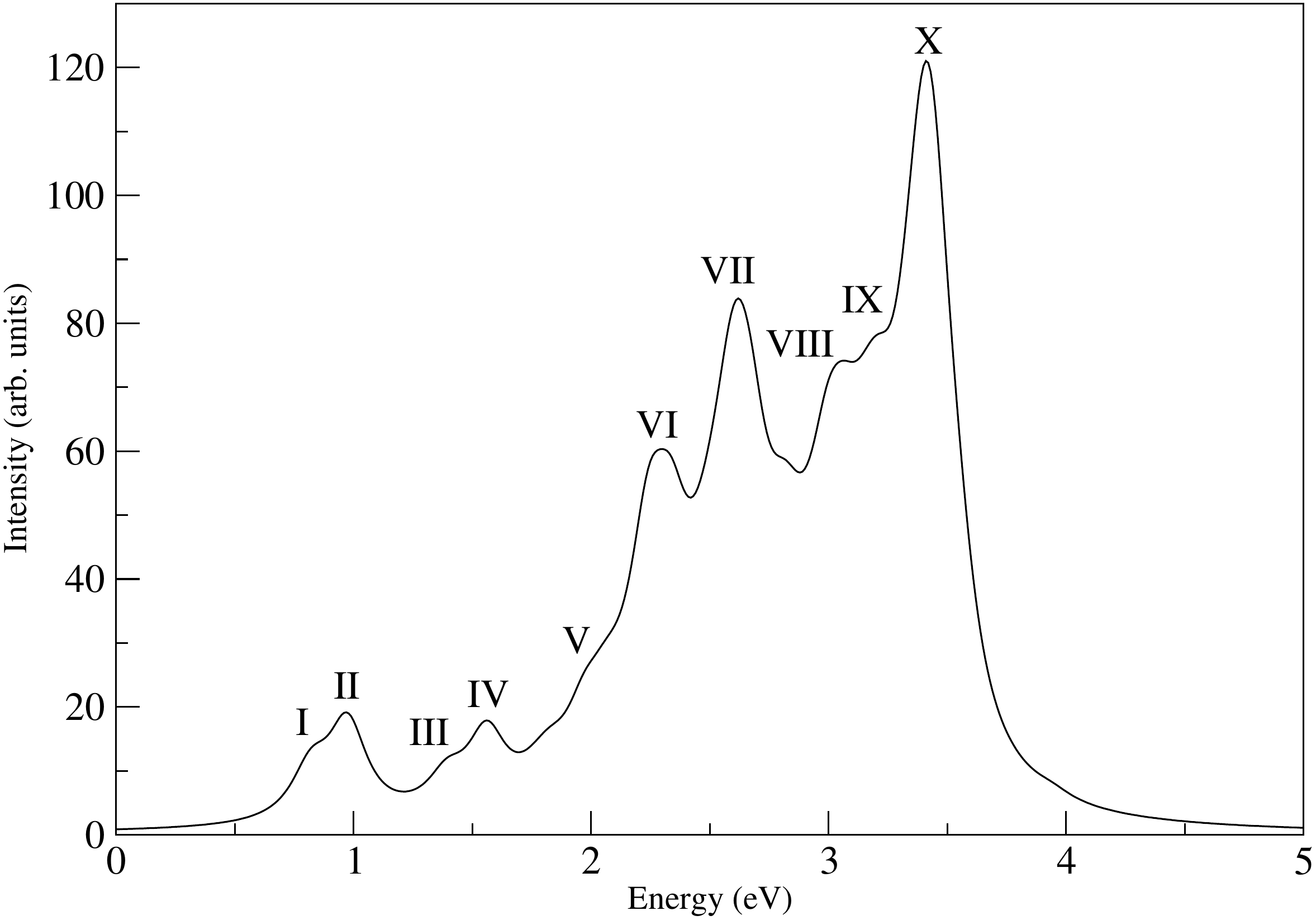}

}

\subfloat[\label{fig:GQD-DV1-585_Singlet}]{\includegraphics[scale=0.25]{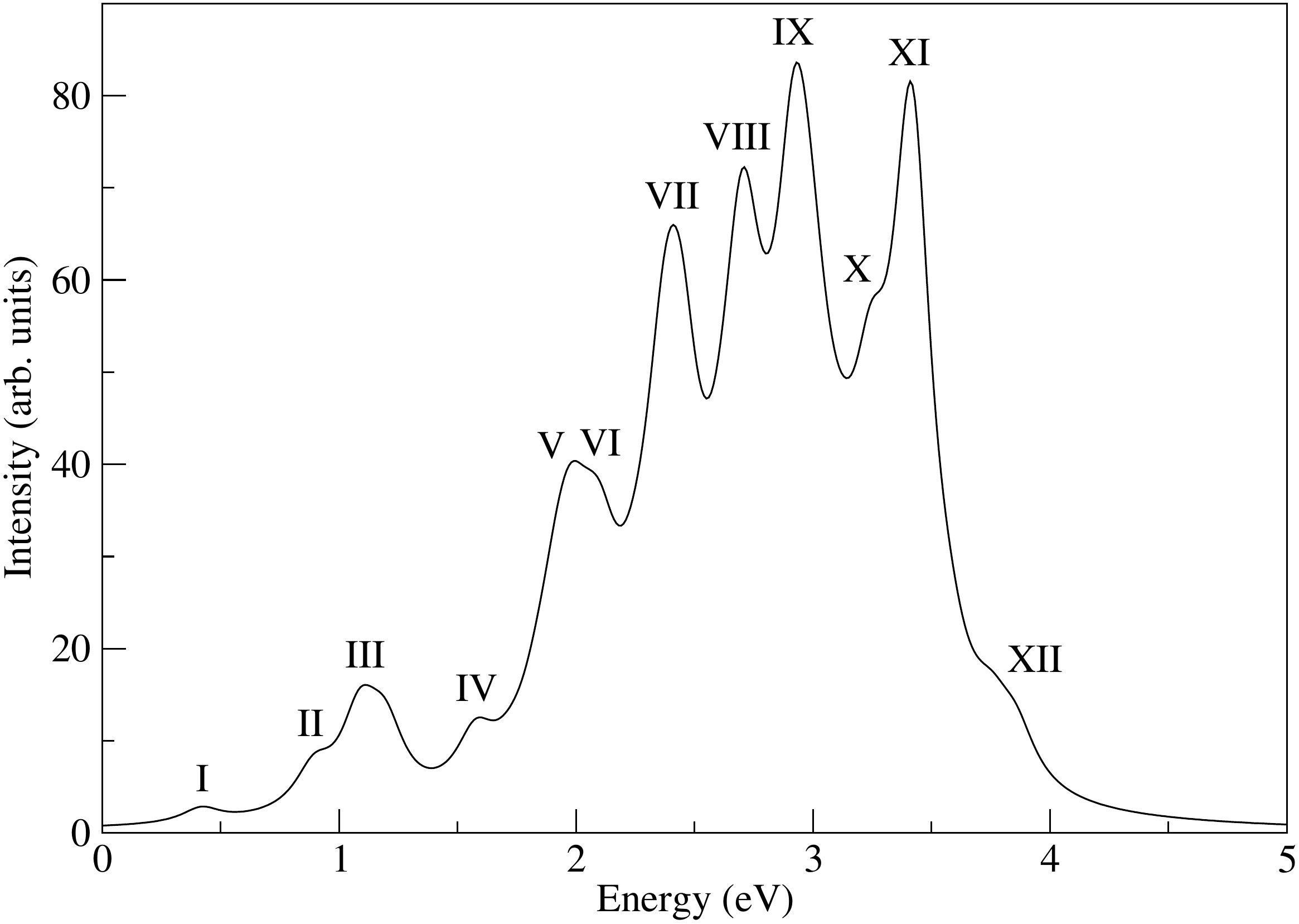}

}

\subfloat[\label{fig:GQD-DV2_Singlet}]{\includegraphics[scale=0.25]{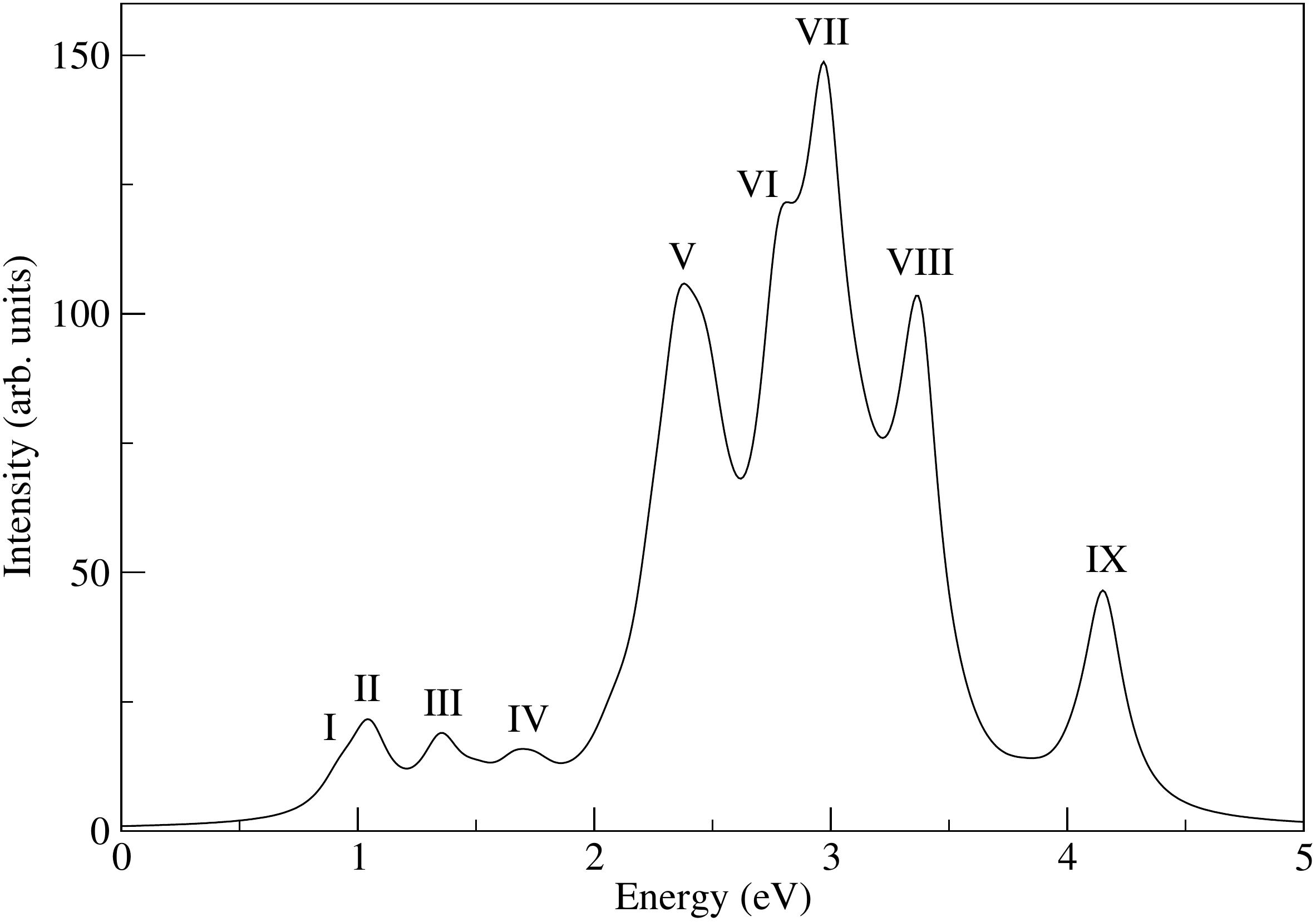}

}

\subfloat[\label{fig:GQD-DV2-594_Singlet}]{\includegraphics[scale=0.25]{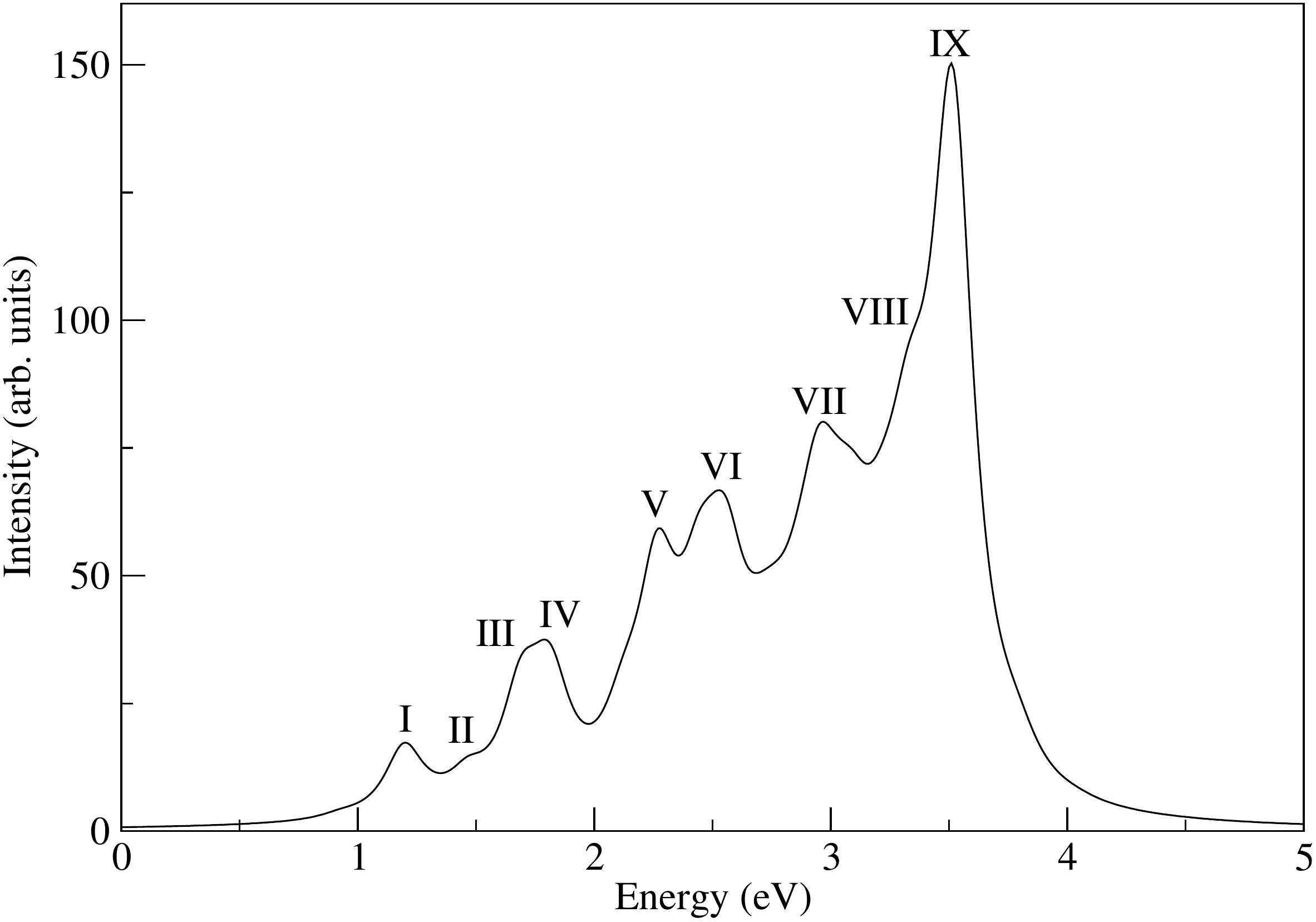}

}\caption{Singlet optical \label{fig:Singlet-optical-absorption}absorption
spectra of (a) GQD-DV1, (b) GQD-DV1-585, (c) GQD-DV2 and (d) GQD-DV2-594
calculated by adopting the PPP-CI approach. These spectra are broadened
with a uniform line width of 0.1 eV.}
\end{figure}

A detailed investigation of the \textcolor{black}{PPP-CI} singlet
absorption spectra (Figs. \ref{fig:Singlet-optical-absorption}(a-d))
and quantitative data of the corresponding excited states (Tables
S3-S6 \cite{Supp_Info}) reveals the following characteristics: (i)
The first peaks of the unrelaxed structures GQD-DV1 and GQD-DV2, are
strongly dominated by transition of a single electron from H to L+1
level, denoted by $|H\rightarrow L+1\rangle$, henceforth. However,
when the SW-defected GQDs with DV undergo relaxation, the first peak
of GQD-DV2-594 is primarily due to double excitation $|H\rightarrow L;H\rightarrow L\rangle$,
while for the case of GQD-DV1-585, this peak is mainly composed of
the single excitation $|H\rightarrow L+1\rangle$, along with substantial
contribution from the doubly excited wave-function $|H\rightarrow L;H\rightarrow L+1\rangle$.
(ii) The singly-excited electronic configuration $|H\rightarrow L\rangle$
is primarily responsible for the second absorption peak of GQD-DV1,
GQD-DV1-585 and GQD-DV2. However, in case of GQD-DV2-594, this excitation
leads to an extremely small hump at energy 0.93 eV, which is hardly
perceptible in the optical absorption spectrum. These results are
primarily a consequence of electron-electron correlations coupled
with broken particle-hole symmetry due to DV and SW-type imperfections
in these GQDs. (iii) The second peak of GQD-DV1 and GQD-DV2, dominantly
due to $|H\rightarrow L\rangle$ transition, has mixed $x-y$ polarization
with higher magnitude of the transition dipole moment along the y
axis, while this peak gets predominantly x-polarized for GQD-DV1-585.
(iv) The position of the most intense (MI) peak is in the UV range
for GQD-DV1 while it shifts to the violet range for its relaxed configuration.
This is in stark contrast to the reverse trend demonstrated by SW-defected
GQDs having divacancies at the reczag edge, wherein the MI peak shifts
from violet to UV range on relaxation. In addition, the magnitudes
of these red/blue-shifts are quite comparable to each other. (v) The
strength of the most intense peak of GQD-DV1 decreases significantly
after relaxation, while no such trend is observed for GQD-DV2. (vi)
The optical absorption peak profile (peak pattern as well as peak
intensity) is drastically altered when the DV-defected GQDs undergo
relaxation. 

\begin{figure}
\subfloat[\label{fig:GQD-DV1_Triplet}]{\includegraphics[scale=0.25]{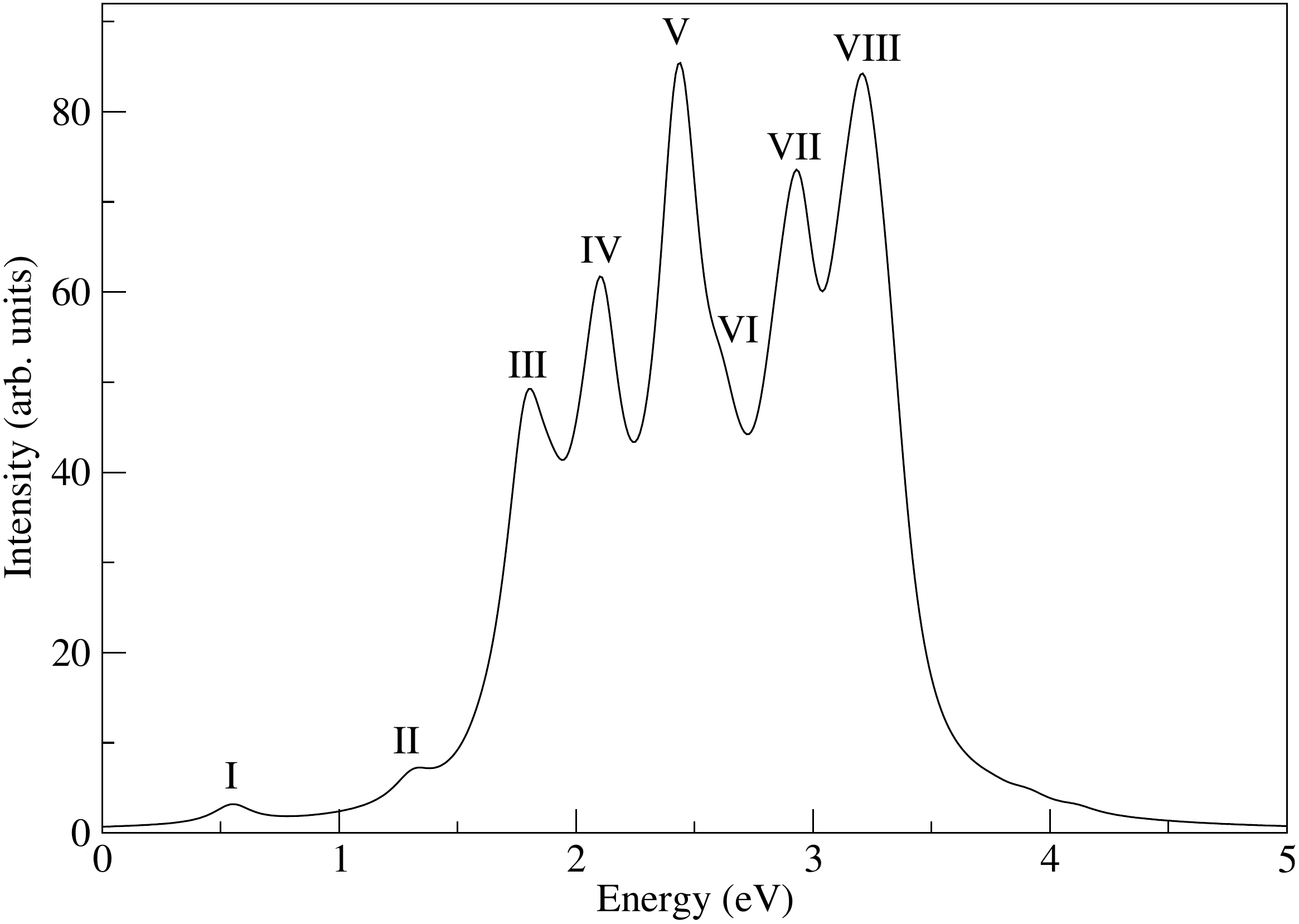}

}

\subfloat[\label{fig:GQD-DV1-585_Triplet}]{\includegraphics[scale=0.25]{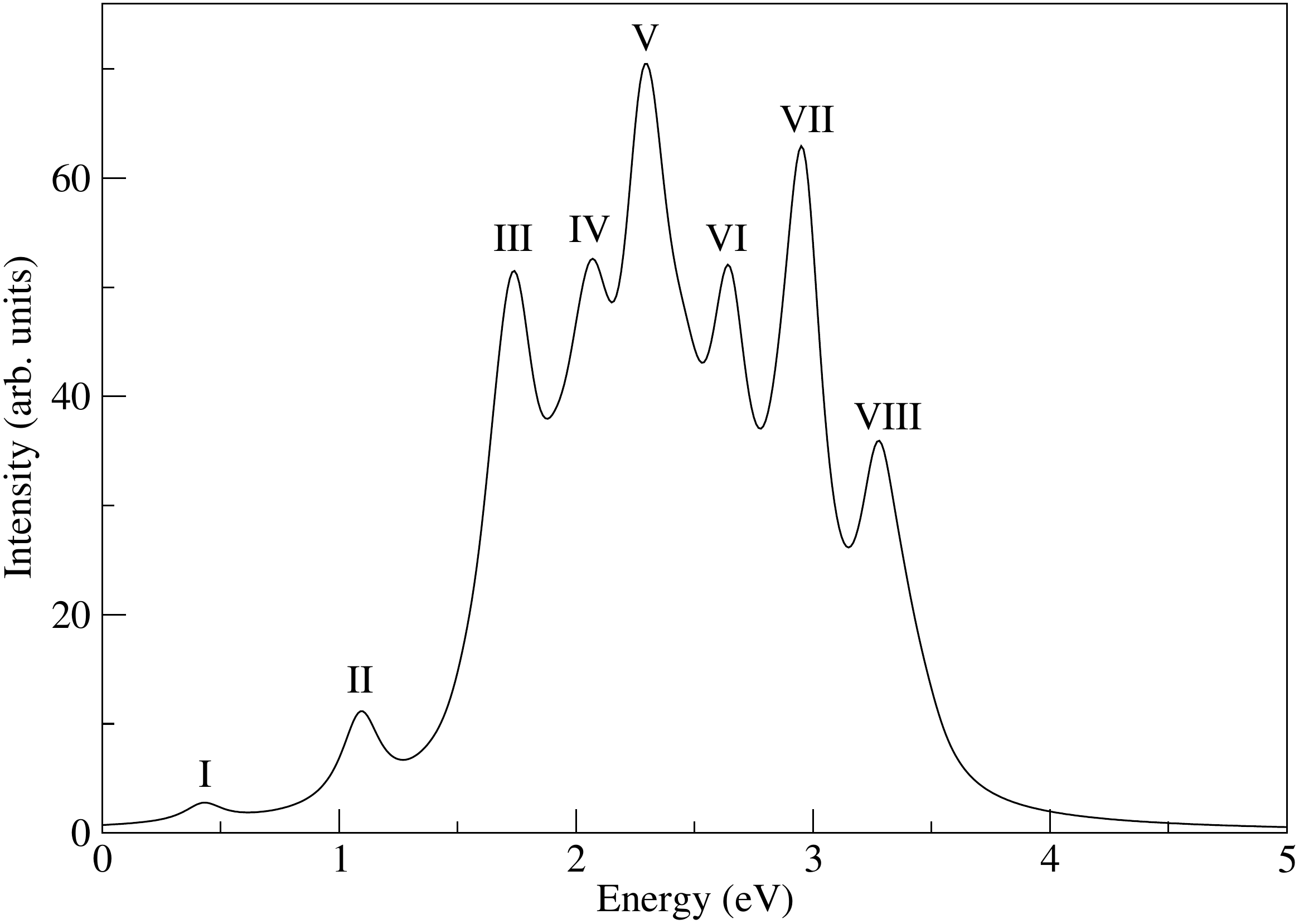}

}

\subfloat[\label{fig:GQD-DV2_Triplet}]{\includegraphics[scale=0.25]{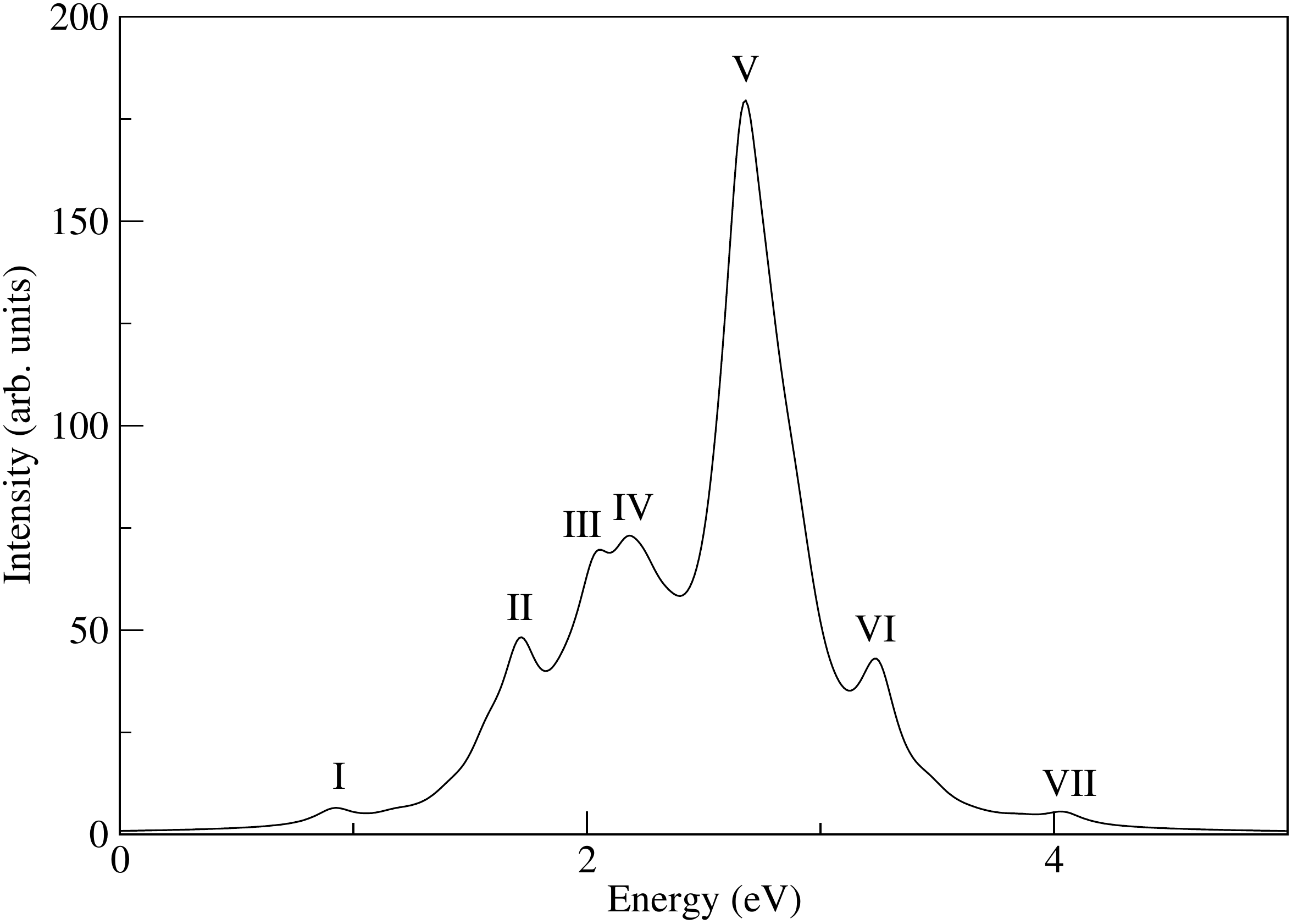}

}

\subfloat[\label{fig:GQD-DV2-594_Triplet}]{\includegraphics[scale=0.25]{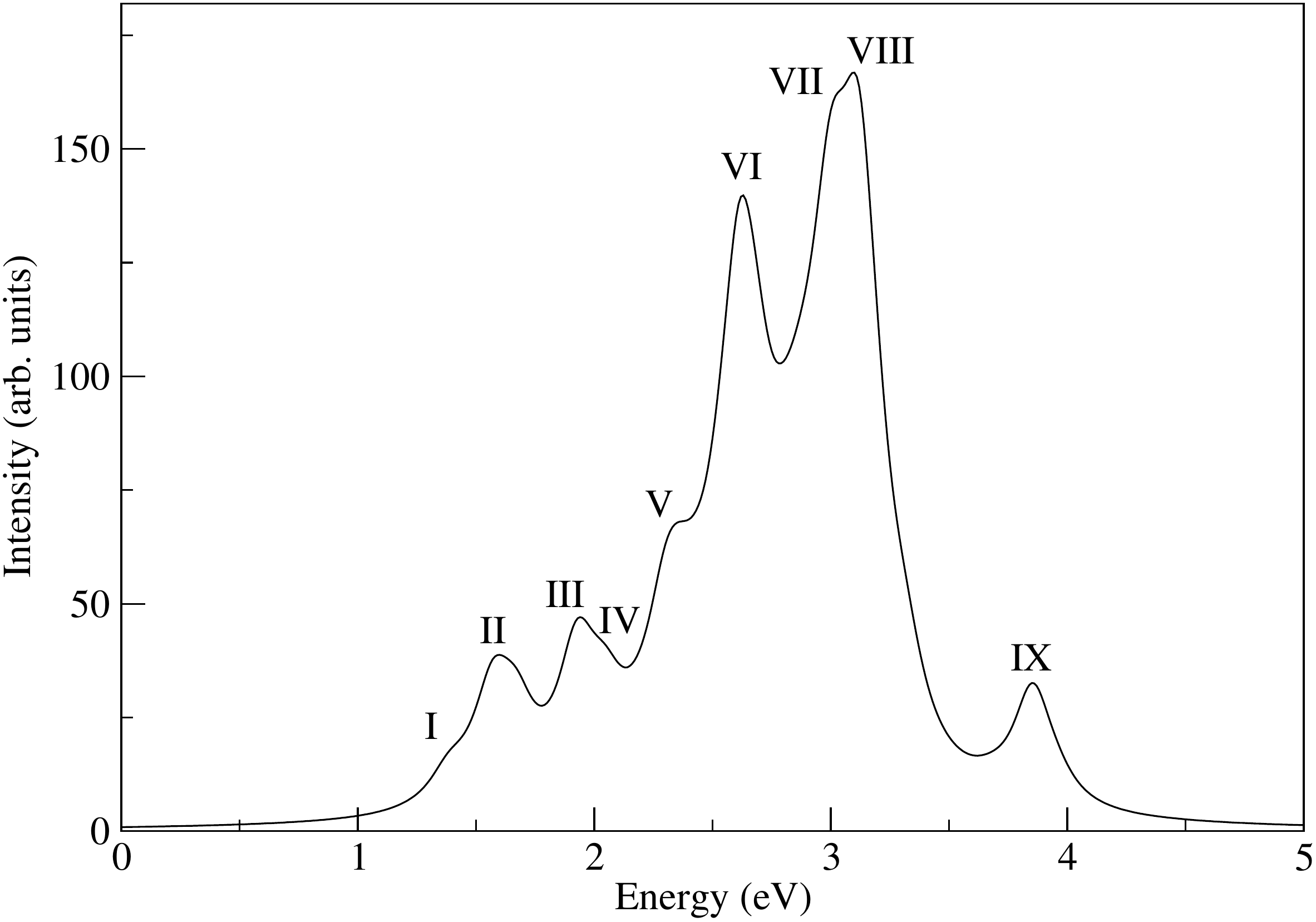}

}\caption{Triplet optical \label{fig:Triplet-optical-absorption}absorption
spectra of (a) GQD-DV1, (b) GQD-DV1-585, (c) GQD-DV2 and (d) GQD-DV2-594
calculated by adopting the PPP-CI approach. These spectra are broadened
with a uniform line width of 0.1 eV. The peak positions represent
the excitation energies of the higher triplet states with respect
to the lowest triplet state. }

\end{figure}

In case of the computed \textcolor{black}{PPP-CI} triplet one-photon
absorption spectra, the peak positions represent the excitation energies
of the higher triplet states with respect to the lowest triplet state.
A detailed examination of this spectra (Figs. \ref{fig:Triplet-optical-absorption}(a-d))
and quantitative data of the corresponding excited states (Tables
S7-S10 \cite{Supp_Info}) elucidates the following features: (i) The
entire triplet spectrum of GQD-DV1, GQD-DV1-585 and GQD-DV2 is red-shifted
while that of GQD-DV2-594 is marginally blue-shifted with respect
to their corresponding singlet spectrum. \textcolor{black}{However,
the MI peak of the triplet absorption spectrum is always red-shifted
compared to its singlet counterpart}. (ii) The MI peak is red/blue-shifted
when SW-defected GQDs with DV imperfections at the centre/reczag edge
undergo relaxation, in agreement with the behaviour exhibited by their
corresponding singlet spectra. The MI peak is in the green band for
both the unrelaxed and relaxed SW-defected GQDs with DVs at the core.
However, the position of MI peak is shifted from blue to violet range
when GQD-DV2 relaxes to GQD-DV2-594 configuration. The magnitude of
the computed blue-shift (0.45 eV) is much higher compared to the red-shift
(0.14 eV) for the triplet spectra of these GQDs. This trend is distinctly
different from the comparable values of the red and blue-shift (0.47
and 0.54 eV, respectively) of the singlet MI peak. (iii) The intensity
of the strongest triplet absorption peak is always less/more than
its singlet counterpart for both relaxed and unrelaxed SW-defected
GQDs with DV imperfections at the centre/reczag edge. Further, the
strength of the triplet MI peak always reduces when GQD-DV1 and GQD-DV2
systems are relaxed, with the reduction being more evident for GQD-DV1
geometry. (iv) The number of absorption peaks in the IR and UV band
is always less in the triplet spectra compared to their corresponding
singlet spectra for all our studied GQDs. However, the number of triplet
optical peaks in the visible energy range increases for GQD-DV1, GQD-DV2
and GQD-DV2-594, while it is equal to the number of singlet peaks
in this energy domain for GQD-DV1-584 configuration. (v) The number
of triplet excited states with major contributions from double excitations
or having significant mixing of many singly excited configurations
is more than the number of singlet excited states with such attributes.
This emphasizes that the strength of electron-electron correlations
is greater for triplet manifold in comparison to singlet states.

\textcolor{black}{A comparison of the MI peak energies of the singlet
absorption spectra of GQD-DV1-585 and GQD-DV2-594, obtained from TDDFT
and PPP-CI approaches (Table \ref{tab:Comparison-of-MI}), reveal
that the TDDFT values are in excellent agreement with the PPP-CI results.
In contrast, there is a large discrepancy between the MI peak energies
of the triplet absorption spectra, computed from TDDFT and PPP-CI
methodologies, with the variation being more pronounced for GQD-DV2-594.
This deviation is due to the fact that TDDFT which is analogous to
CI-singles (CIS) methodology considers only single particle-hole excitations
and therefore, cannot accurately describe the triplet states where
electron correlation effects are highly dominant. In addition, the
TDDFT values of the MI peak of the triplet absorption spectrum is
always red-shifted compared to its singlet counterpart, in agreement
with the PPP-CI results. Hence, our computations demonstrate that
the PPP-CI approach is more effective than TDDFT methodology in predicting
the trends of the singlet and triplet absorption spectrum of these
defected systems.}

\begin{table}
\caption{Comparison of \label{tab:Comparison-of-MI}MI peak energies of the
singlet and triplet absorption spectra of GQD-DV1-585 and GQD-DV2-594,
obtained from TDDFT and PPP-CI approaches.}

\begin{tabular}{|c|c|c|c|c|}
\hline 
 & \multicolumn{4}{c|}{MI peak energy (eV) }\tabularnewline
\cline{2-5} 
System & \multicolumn{2}{c|}{Singlet Absorption Spectrum} & \multicolumn{2}{c|}{Triplet Absorption Spectrum }\tabularnewline
\cline{2-5} 
 & TDDFT & PPP-CI & TDDFT & PPP-CI\tabularnewline
\hline 
GQD-DV1-585 & 3.01 & 2.93 & 2.66 & 2.28\tabularnewline
\hline 
GQD-DV2-594 & 3.53 & 3.52 & 2.56 & 3.09, 3.12\tabularnewline
\hline 
\end{tabular}

\end{table}

\section{Conclusions}

In summary, we have successfully captured the signatures of structural
relaxation in governing the electronic and optical properties of SW-defected
GQDs with divacancy by performing extensive electron-correlated computations
based on the PPP Hamiltonian \textcolor{black}{and TDDFT approach}.
It is observed that the geometry of the relaxed systems is critically
dependent on the position of the divacancy in these structur\textcolor{black}{es.
Our results conclusively show that structural relaxation switches
the electronic ground-state configuration of these systems from low-spin
singlet to high-spin triplet state. The thermal stability of this
high-spin triplet ground state of the relaxed systems at and above
the room temperature significantly augments the applicability of these
materials in designing spintronic devices. Notably, our computations
reveal that the energetic ordering of the singlet and triplet states
of the relaxed systems facilitates the possibility of achieving both
prompt as well as delayed fluorescence of different wavelengths and
the tunability range of the emission wavelengths is determined by
the position of divacancy in these SW-defected GQDs. Our fi}ndings
signify that the entire singlet and triplet optical absorption profile
(peak pattern as well as intensity) is substantially altered due to
structural relaxation. Specifically, the nature of shift exhibited
by the most intense absorption peak of the relaxed GQDs with respect
to their unrelaxed counterpart depends on the location of the divacancy.
The magnitude of this shift is determined by the spin multiplicity
of the energy states. \textcolor{black}{The spin multiplicity also
has distinct imprints on the entire absorption peak profile of these
systems.} \textcolor{black}{Also, all these marked features of the
optical absorption spectrum can be more accurately described by the
PPP-CI methodology as compared to the TDDFT approach.} In conclusion,
our results provide new insights to the novel concept of exploiting
SW-defected GQDs with divacancy for designing highly efficient optoelectronic
and spintronic devices. 
\begin{suppinfo}
The \textcolor{black}{coordinates of the carbon and hydrogen atoms},
dimensions of CI matrices considered for computing the linear optical
absorption spectra of singlet and triplet spin multiplicity of GQD-DV1,
GQD-DV1-585, GQD-DV2 and GQD-DV2-594 systems, energies, transition
dipole moments and the configurations contributing dominantly to the
many-particle wave functions of the excited states primarily responsible
for the absorption peaks in the PPP-CI singlet and triplet spectra
of GQD-DV1, GQD-DV1-585, GQD-DV2 and GQD-DV2-594 configurations are
included in the supporting information.
\end{suppinfo}
\addcontentsline{toc}{section}{\refname}\bibliography{DV_Reczag}

\providecommand{\latin}[1]{#1}
\makeatletter
\providecommand{\doi}
  {\begingroup\let\do\@makeother\dospecials
  \catcode`\{=1 \catcode`\}=2 \doi@aux}
\providecommand{\doi@aux}[1]{\endgroup\texttt{#1}}
\makeatother
\providecommand*\mcitethebibliography{\thebibliography}
\csname @ifundefined\endcsname{endmcitethebibliography}
  {\let\endmcitethebibliography\endthebibliography}{}
\begin{mcitethebibliography}{28}
\providecommand*\natexlab[1]{#1}
\providecommand*\mciteSetBstSublistMode[1]{}
\providecommand*\mciteSetBstMaxWidthForm[2]{}
\providecommand*\mciteBstWouldAddEndPuncttrue
  {\def\EndOfBibitem{\unskip.}}
\providecommand*\mciteBstWouldAddEndPunctfalse
  {\let\EndOfBibitem\relax}
\providecommand*\mciteSetBstMidEndSepPunct[3]{}
\providecommand*\mciteSetBstSublistLabelBeginEnd[3]{}
\providecommand*\EndOfBibitem{}
\mciteSetBstSublistMode{f}
\mciteSetBstMaxWidthForm{subitem}{(\alph{mcitesubitemcount})}
\mciteSetBstSublistLabelBeginEnd
  {\mcitemaxwidthsubitemform\space}
  {\relax}
  {\relax}

\bibitem[Hirata(2017)]{Hirata_Phosphorescence}
Hirata,~S. Recent Advances in Materials with Room-Temperature Phosphorescence:
  Photophysics for Triplet Exciton Stabilization. \emph{Advanced Optical
  Materials} \textbf{2017}, \emph{5}, 1700116\relax
\mciteBstWouldAddEndPuncttrue
\mciteSetBstMidEndSepPunct{\mcitedefaultmidpunct}
{\mcitedefaultendpunct}{\mcitedefaultseppunct}\relax
\EndOfBibitem
\bibitem[Hashimoto \latin{et~al.}(2004)Hashimoto, Suenaga, Gloter, Urita, and
  Iijima]{Hashimoto_Nature_2004}
Hashimoto,~A.; Suenaga,~K.; Gloter,~A.; Urita,~K.; Iijima,~S. Direct evidence
  for atomic defects in graphene layers. \emph{Nature} \textbf{2004},
  \emph{430}, 870\relax
\mciteBstWouldAddEndPuncttrue
\mciteSetBstMidEndSepPunct{\mcitedefaultmidpunct}
{\mcitedefaultendpunct}{\mcitedefaultseppunct}\relax
\EndOfBibitem
\bibitem[Meyer \latin{et~al.}(2008)Meyer, Kisielowski, Erni, Rossell, Crommie,
  and Zettl]{Meyer_Nano_Letters_2008}
Meyer,~J.~C.; Kisielowski,~C.; Erni,~R.; Rossell,~M.~D.; Crommie,~M.~F.;
  Zettl,~A. Direct Imaging of Lattice Atoms and Topological Defects in Graphene
  Membranes. \emph{Nano Letters} \textbf{2008}, \emph{8}, 3582--3586\relax
\mciteBstWouldAddEndPuncttrue
\mciteSetBstMidEndSepPunct{\mcitedefaultmidpunct}
{\mcitedefaultendpunct}{\mcitedefaultseppunct}\relax
\EndOfBibitem
\bibitem[Cretu \latin{et~al.}(2010)Cretu, Krasheninnikov, Rodriguez-Manzo, Sun,
  Nieminen, and Banhart]{Cretu_PhysRevLett.105.196102}
Cretu,~O.; Krasheninnikov,~A.~V.; Rodriguez-Manzo,~J.~A.; Sun,~L.;
  Nieminen,~R.~M.; Banhart,~F. Migration and Localization of Metal Atoms on
  Strained Graphene. \emph{Phys. Rev. Lett.} \textbf{2010}, \emph{105},
  196102\relax
\mciteBstWouldAddEndPuncttrue
\mciteSetBstMidEndSepPunct{\mcitedefaultmidpunct}
{\mcitedefaultendpunct}{\mcitedefaultseppunct}\relax
\EndOfBibitem
\bibitem[Girit \latin{et~al.}(2009)Girit, Meyer, Erni, Rossell, Kisielowski,
  Yang, Park, Crommie, Cohen, Louie, and Zettl]{Girit_Science_2009}
Girit,~C.~O.; Meyer,~J.~C.; Erni,~R.; Rossell,~M.~D.; Kisielowski,~C.;
  Yang,~L.; Park,~C.-H.; Crommie,~M.~F.; Cohen,~M.~L.; Louie,~S.~G.; Zettl,~A.
  Graphene at the Edge: Stability and Dynamics. \emph{Science} \textbf{2009},
  \emph{323}, 1705--1708\relax
\mciteBstWouldAddEndPuncttrue
\mciteSetBstMidEndSepPunct{\mcitedefaultmidpunct}
{\mcitedefaultendpunct}{\mcitedefaultseppunct}\relax
\EndOfBibitem
\bibitem[Kim \latin{et~al.}(2011)Kim, Ihm, Yoon, and Lee]{Kim_PRB_2011}
Kim,~Y.; Ihm,~J.; Yoon,~E.; Lee,~G.-D. Dynamics and stability of divacancy
  defects in graphene. \emph{Phys. Rev. B} \textbf{2011}, \emph{84},
  075445\relax
\mciteBstWouldAddEndPuncttrue
\mciteSetBstMidEndSepPunct{\mcitedefaultmidpunct}
{\mcitedefaultendpunct}{\mcitedefaultseppunct}\relax
\EndOfBibitem
\bibitem[Lee \latin{et~al.}(2005)Lee, Wang, Yoon, Hwang, Kim, and
  Ho]{Lee_PRL_2005}
Lee,~G.-D.; Wang,~C.~Z.; Yoon,~E.; Hwang,~N.-M.; Kim,~D.-Y.; Ho,~K.~M.
  Diffusion, Coalescence, and Reconstruction of Vacancy Defects in Graphene
  Layers. \emph{Phys. Rev. Lett.} \textbf{2005}, \emph{95}, 205501\relax
\mciteBstWouldAddEndPuncttrue
\mciteSetBstMidEndSepPunct{\mcitedefaultmidpunct}
{\mcitedefaultendpunct}{\mcitedefaultseppunct}\relax
\EndOfBibitem
\bibitem[Leyssale and Vignoles(2014)Leyssale, and Vignoles]{Leyssale_jp501028n}
Leyssale,~J.-M.; Vignoles,~G.~L. A Large-Scale Molecular Dynamics Study of the
  Divacancy Defect in Graphene. \emph{The Journal of Physical Chemistry C}
  \textbf{2014}, \emph{118}, 8200--8216\relax
\mciteBstWouldAddEndPuncttrue
\mciteSetBstMidEndSepPunct{\mcitedefaultmidpunct}
{\mcitedefaultendpunct}{\mcitedefaultseppunct}\relax
\EndOfBibitem
\bibitem[Krasheninnikov \latin{et~al.}(2006)Krasheninnikov, Lehtinen, Foster,
  and Nieminen]{Krasheninnikov_2006_CPL}
Krasheninnikov,~A.; Lehtinen,~P.; Foster,~A.; Nieminen,~R. Bending the rules:
  Contrasting vacancy energetics and migration in graphite and carbon
  nanotubes. \emph{Chemical Physics Letters} \textbf{2006}, \emph{418},
  132--136\relax
\mciteBstWouldAddEndPuncttrue
\mciteSetBstMidEndSepPunct{\mcitedefaultmidpunct}
{\mcitedefaultendpunct}{\mcitedefaultseppunct}\relax
\EndOfBibitem
\bibitem[Kotakoski \latin{et~al.}(2006)Kotakoski, Krasheninnikov, and
  Nordlund]{Kotakoski_PRB_2006}
Kotakoski,~J.; Krasheninnikov,~A.~V.; Nordlund,~K. Energetics, structure, and
  long-range interaction of vacancy-type defects in carbon nanotubes: Atomistic
  simulations. \emph{Phys. Rev. B} \textbf{2006}, \emph{74}, 245420\relax
\mciteBstWouldAddEndPuncttrue
\mciteSetBstMidEndSepPunct{\mcitedefaultmidpunct}
{\mcitedefaultendpunct}{\mcitedefaultseppunct}\relax
\EndOfBibitem
\bibitem[El-Barbary \latin{et~al.}(2003)El-Barbary, Telling, Ewels, Heggie, and
  Briddon]{El_barbary_PRB_2003}
El-Barbary,~A.~A.; Telling,~R.~H.; Ewels,~C.~P.; Heggie,~M.~I.; Briddon,~P.~R.
  Structure and energetics of the vacancy in graphite. \emph{Phys. Rev. B}
  \textbf{2003}, \emph{68}, 144107\relax
\mciteBstWouldAddEndPuncttrue
\mciteSetBstMidEndSepPunct{\mcitedefaultmidpunct}
{\mcitedefaultendpunct}{\mcitedefaultseppunct}\relax
\EndOfBibitem
\bibitem[Sahan and Berber(2019)Sahan, and Berber]{Sahan_PhysicaE_2019}
Sahan,~Z.; Berber,~S. Divacancy in graphene nano-ribbons. \emph{Physica E:
  Low-dimensional Systems and Nanostructures} \textbf{2019}, \emph{106},
  239--249\relax
\mciteBstWouldAddEndPuncttrue
\mciteSetBstMidEndSepPunct{\mcitedefaultmidpunct}
{\mcitedefaultendpunct}{\mcitedefaultseppunct}\relax
\EndOfBibitem
\bibitem[Liu and Chen(2017)Liu, and Chen]{Liu_Hindawi_2017}
Liu,~L.; Chen,~S. Geometries and Electronic States of Divacancy Defect in
  Finite-Size Hexagonal Graphene Flakes. \emph{Journal of Chemistry}
  \textbf{2017}, \emph{2017}, 8491264\relax
\mciteBstWouldAddEndPuncttrue
\mciteSetBstMidEndSepPunct{\mcitedefaultmidpunct}
{\mcitedefaultendpunct}{\mcitedefaultseppunct}\relax
\EndOfBibitem
\bibitem[Yumura \latin{et~al.}(2012)Yumura, Awano, Kobayashi, and
  Yamabe]{Yumura_Molecules_2012}
Yumura,~T.; Awano,~T.; Kobayashi,~H.; Yamabe,~T. Platinum Clusters on
  Vacancy-Type Defects of Nanometer-Sized Graphene Patches. \emph{Molecules}
  \textbf{2012}, \emph{17}, 7941--7960\relax
\mciteBstWouldAddEndPuncttrue
\mciteSetBstMidEndSepPunct{\mcitedefaultmidpunct}
{\mcitedefaultendpunct}{\mcitedefaultseppunct}\relax
\EndOfBibitem
\bibitem[Frisch \latin{et~al.}(2016)Frisch, Trucks, Schlegel, Scuseria, Robb,
  Cheeseman, Scalmani, Barone, Petersson, Nakatsuji, and et~al.]{g16}
Frisch,~M.~J.; Trucks,~G.~W.; Schlegel,~H.~B.; Scuseria,~G.~E.; Robb,~M.~A.;
  Cheeseman,~J.~R.; Scalmani,~G.; Barone,~V.; Petersson,~G.~A.; Nakatsuji,~H.;
  et~al., Gaussian 16 {R}evision {C}.01. 2016; Gaussian Inc. Wallingford
  CT\relax
\mciteBstWouldAddEndPuncttrue
\mciteSetBstMidEndSepPunct{\mcitedefaultmidpunct}
{\mcitedefaultendpunct}{\mcitedefaultseppunct}\relax
\EndOfBibitem
\bibitem[Pople(1953)]{ppp-pople}
Pople,~J.~A. Electron interaction in unsaturated hydrocarbons. \emph{Trans.
  Faraday Soc.} \textbf{1953}, \emph{49}, 1375--1385\relax
\mciteBstWouldAddEndPuncttrue
\mciteSetBstMidEndSepPunct{\mcitedefaultmidpunct}
{\mcitedefaultendpunct}{\mcitedefaultseppunct}\relax
\EndOfBibitem
\bibitem[Pariser and Parr(1953)Pariser, and Parr]{ppp-pariser-parr}
Pariser,~R.; Parr,~R.~G. A SemiEmpirical Theory of the Electronic Spectra and
  Electronic Structure of Complex Unsaturated Molecules. II. \emph{J. Chem.
  Phys.} \textbf{1953}, \emph{21}, 767--776\relax
\mciteBstWouldAddEndPuncttrue
\mciteSetBstMidEndSepPunct{\mcitedefaultmidpunct}
{\mcitedefaultendpunct}{\mcitedefaultseppunct}\relax
\EndOfBibitem
\bibitem[Basak \latin{et~al.}(2021)Basak, Basak, and
  Shukla]{Tista_PhysRevB.103.235420}
Basak,~T.; Basak,~T.; Shukla,~A. Graphene quantum dots with a Stone-Wales
  defect as a topologically tunable platform for visible-light harvesting.
  \emph{Phys. Rev. B} \textbf{2021}, \emph{103}, 235420\relax
\mciteBstWouldAddEndPuncttrue
\mciteSetBstMidEndSepPunct{\mcitedefaultmidpunct}
{\mcitedefaultendpunct}{\mcitedefaultseppunct}\relax
\EndOfBibitem
\bibitem[Ohno(1964)]{Theor.chim.act.2Ohno}
Ohno,~K. Some remarks on the Pariser-Parr-Pople method. \emph{Theoretica
  chimica acta} \textbf{1964}, \emph{2}, 219--227\relax
\mciteBstWouldAddEndPuncttrue
\mciteSetBstMidEndSepPunct{\mcitedefaultmidpunct}
{\mcitedefaultendpunct}{\mcitedefaultseppunct}\relax
\EndOfBibitem
\bibitem[Basak \latin{et~al.}(2018)Basak, Basak, and
  Shukla]{PhysRevB.98.035401}
Basak,~T.; Basak,~T.; Shukla,~A. Electron correlation effects and two-photon
  absorption in diamond-shaped graphene quantum dots. \emph{Phys. Rev. B}
  \textbf{2018}, \emph{98}, 035401\relax
\mciteBstWouldAddEndPuncttrue
\mciteSetBstMidEndSepPunct{\mcitedefaultmidpunct}
{\mcitedefaultendpunct}{\mcitedefaultseppunct}\relax
\EndOfBibitem
\bibitem[Basak and Shukla(2016)Basak, and Shukla]{Tista_PRB93}
Basak,~T.; Shukla,~A. Optical signatures of electric-field-driven magnetic
  phase transitions in graphene quantum dots. \emph{Phys. Rev. B}
  \textbf{2016}, \emph{93}, 235432\relax
\mciteBstWouldAddEndPuncttrue
\mciteSetBstMidEndSepPunct{\mcitedefaultmidpunct}
{\mcitedefaultendpunct}{\mcitedefaultseppunct}\relax
\EndOfBibitem
\bibitem[Basak \latin{et~al.}(2015)Basak, Chakraborty, and Shukla]{Tista_PRB92}
Basak,~T.; Chakraborty,~H.; Shukla,~A. Theory of linear optical absorption in
  diamond-shaped graphene quantum dots. \emph{Phys. Rev. B} \textbf{2015},
  \emph{92}, 205404\relax
\mciteBstWouldAddEndPuncttrue
\mciteSetBstMidEndSepPunct{\mcitedefaultmidpunct}
{\mcitedefaultendpunct}{\mcitedefaultseppunct}\relax
\EndOfBibitem
\bibitem[Basak and Basak(2020)Basak, and
  Basak]{Tista_Basak_Mat_Today_Proc_2020}
Basak,~T.; Basak,~T. Electro-absorption spectra of magnetic states of diamond
  shaped graphene quantum dots. \emph{Materials Today: Proceedings}
  \textbf{2020}, \emph{26}, 2058 -- 2061, 10th International Conference of
  Materials Processing and Characterization\relax
\mciteBstWouldAddEndPuncttrue
\mciteSetBstMidEndSepPunct{\mcitedefaultmidpunct}
{\mcitedefaultendpunct}{\mcitedefaultseppunct}\relax
\EndOfBibitem
\bibitem[Basak and Basak(2020)Basak, and
  Basak]{Tushima_Basak_Mat_Today_Proc_2020}
Basak,~T.; Basak,~T. Characterization of magnetic states of graphene quantum
  dots of different shapes by application of electric field. \emph{Materials
  Today: Proceedings} \textbf{2020}, \emph{26}, 2069 -- 2072, 10th
  International Conference of Materials Processing and Characterization\relax
\mciteBstWouldAddEndPuncttrue
\mciteSetBstMidEndSepPunct{\mcitedefaultmidpunct}
{\mcitedefaultendpunct}{\mcitedefaultseppunct}\relax
\EndOfBibitem
\bibitem[Sony and Shukla(2010)Sony, and Shukla]{Sony2010821}
Sony,~P.; Shukla,~A. A general purpose Fortran 90 electronic structure program
  for conjugated systems using Pariser-Parr-Pople model. \emph{Computer Physics
  Communications} \textbf{2010}, \emph{181}, 821 -- 830\relax
\mciteBstWouldAddEndPuncttrue
\mciteSetBstMidEndSepPunct{\mcitedefaultmidpunct}
{\mcitedefaultendpunct}{\mcitedefaultseppunct}\relax
\EndOfBibitem
\bibitem[Sup()]{Supp_Info}
Supporting Information. \relax
\mciteBstWouldAddEndPunctfalse
\mciteSetBstMidEndSepPunct{\mcitedefaultmidpunct}
{}{\mcitedefaultseppunct}\relax
\EndOfBibitem
\bibitem[Hu \latin{et~al.}(2015)Hu, Yao, Yang, and Ma]{Hu_RISC}
Hu,~D.; Yao,~L.; Yang,~B.; Ma,~Y. Reverse intersystem crossing from upper
  triplet levels to excited singlet: a hot excition path for organic
  light-emitting diodes. \emph{Philosophical Transactions of the Royal Society
  A: Mathematical, Physical and Engineering Sciences} \textbf{2015},
  \emph{373}, 20140318\relax
\mciteBstWouldAddEndPuncttrue
\mciteSetBstMidEndSepPunct{\mcitedefaultmidpunct}
{\mcitedefaultendpunct}{\mcitedefaultseppunct}\relax
\EndOfBibitem
\end{mcitethebibliography}

\begin{tocentry}
\includegraphics[scale=0.8]{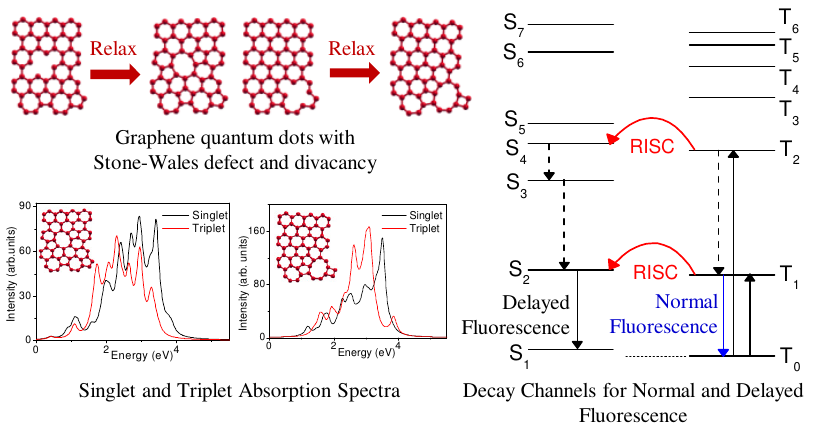}
\end{tocentry}

\end{document}